 \documentclass[aps,pre,reprint,groupedaddress, floatfix, showpacs]{revtex4-1}
 \usepackage{amsmath}
 \usepackage{graphicx}
 \usepackage{bbding}       
 \usepackage{amssymb}      

\begin{document}


\title{Current of interacting particles inside a channel of exponential cavities:
       \\Application of a modified Fick--Jacobs equation}

\author{G. Su\'arez}
\email{gsuarez@mdp.edu.ar}
\author{M. Hoyuelos}
\author{H. M\'artin}
\affiliation{Instituto de Investigaciones F\'isicas de Mar del Plata (IFIMAR -- CONICET)}
\affiliation{Departamento de F\'isica, Facultad de Ciencias Exactas y Naturales,
             Universidad Nacional de Mar del Plata.\\
             De\'an Funes 3350, 7600 Mar del Plata, Argentina}

\date{\today}

\begin{abstract}
The Fick--Jacobs equation has been widely studied, because of its applications in the diffusion and transport of non-interacting particles in narrow channels. It is also known that a modified version of this equation can be used to describe the same system with particles interacting through a hard-core potential. In this work we present a system that can be exactly solved using the Fick--Jacobs equation.
The exact results of the particle concentration profile along the channel $n$, the current, $J$, and the mobility, $\mu$, of particles as a function of an external force are contrasted with Monte Carlo simulations results of non-interacting particles.
For interacting particles the behavior of $n$, $J$ and $\mu$, obtained from the modified Fick--Jacobs equation are in agreement with numerical simulations, where the hard-core interaction is taken into account.
Even more, for interacting particles the modified Fick--Jacobs equation gives comparatively more accurate results of the current difference (when a force is applied in opposite directions) than the exact result for the non-interacting ones.

\end{abstract}


\maketitle

\section{Introduction}
 In the last two decades the diffusion of particles in narrow channels has been the focus of much research \cite{Murphy1972, Batchelor1976, Felderhof1978, Ohtsuki1982, Ohtsuki1983, Lekkerkerker1984, Cichocki1990, Liu1999, Aranovich2005, sevel2006, Henle2008, cheng2008, Burada2009-1, Liu2009, wang2009, Riefler2010, Rubi2010, Borromeo2010, wang2010, Bruna2012, Reguera2012, Ghosh2012, Hofling2013, Malgaretti2013, Li2013, Ai2014, Bruna2014, chen2014, Wang2015, martens2015, wang2015-2, garcia2015}.
Analytically, this problem has been tackled from different angles, however the most accepted description is through the Fick--Jacobs equation, see \cite {jacobs1967, Zwanzig1992, Curado2003, Nobre2004, frank2005, Kalinay2007, Burada2009-2, Martens2011}.

Then, the Fick--Jacobs equation is strictly valid in the regime of very low concentration of diffusing particles in a narrow channel of varying cross section, when the interactions are negligible. When the concentration is not small the interaction plays a crucial role and the behavior of the system can be strongly modified. This has been shown for the case of the hard--core interaction, in \cite{Suarez2013}. Very recently, a generalization of the Fick--Jacobs equation that takes into account this interaction has been introduced. Due to the hard-core interaction this new equation is nonlinear when the particles are dragged along the channel by an external force.

It has been verified that this modified Fick--Jacobs equation successfully reproduces the concentration of particles along the channel, in the stationary state, when contrasted with simulation results \cite{Suarez2015}.
The aim of the paper is to check the validity of the recently introduced nonlinear Fick--Jacobs equation computing the current and the mobility of particles as a function of the external force.
For the sake of completeness the particle concentration profile along the channel used in the present work will be also analysed. These results are contrasted with numerical simulations of the system. 

The transport along channels composed by asymmetrical cavities has recently attracted a great deal of interest due to its possible application to the separation of particles of similar size (see, for example \cite{Reguera2012}). We use an asymmetrical cavity where the cross section is determined by the combination of two exponential functions. 
The purpose of using a cavity with exponential cross section is that for this cavity the original Fick--Jacobs equation for non-interacting particles can be exactly solved. One of the objectives is to compare the accuracy of the modified Fick--Jacobs equation for interacting particles with the accuracy of the original Fick--Jacobs equation, contrasting their results with the corresponding Monte Carlo (MC) simulations.

This article is organized as follows. 
The details of the theoretical description are given in Section~\ref{sec:teo-sinhc} for non--interacting particles, and in Section~\ref{sec:teo-conhc} for interacting ones. 
In Section~\ref{sec:simulations}, we explain the method used to obtain the numerical simulations. All the results and the main analysis of the present work is presented in Section~\ref{sec:results}. Some final remarks are exposed in Section \ref{sec:conclusion}.

\section{Exact solution to the Fick--Jacobs equation inside an exponential cavity}
\label{sec:teo-sinhc} 

As can be seen from \cite{Zwanzig1992} and \cite[p. 68]{jacobs1967} the equation that describes the transport process of non-interacting Brownian particles inside a tube of variable width, and bounded to an external force $F$, is the so called Fick--Jacobs equation. It can be written in terms of the transversally integrated current $J$ given by
\begin{equation}
\frac{J}{D} = - A \frac{\partial n}{\partial x} + F A \beta n \;\;,
\label{eq:FJ-original}
\end{equation}
where $D$ is the diffusion constant, $\beta^{-1} = k_B T$, $A \equiv A(x)$ is the width of the channel at the $x$ position, $n(x)$ is the concentration that is considered constant along the transverse direction of the channel, and $J$ is the total number of particles that cross the transversal area $A$ per unit of time.

\begin{figure}
    \includegraphics[width=\linewidth]{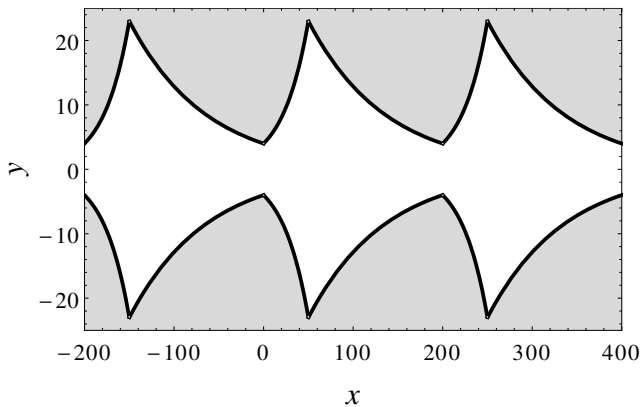}
    \caption{\label{fig:1} Asymmetric cavity described by a combination of exponential functions, see Eq.~(\ref{eq:cavity}). The parameters used are  $A_1 = 4$, $a_1 = 0.035$, $\ell = 50$ and $L = 200$, where the distances are expressed in terms of the lattice spacing $a$ used in the MC simulations.}
\end{figure}

Depending upon the complexity in the shape of the cavity, this equation can be particularly difficult to solve. However in some exceptional cases it is possible to find the exact solution to Eq.~(\ref{eq:FJ-original}). In particular, we considered a chain of identical asymmetric cavities, see Fig.~(\ref{fig:1}). Each of these two--dimensional cavities is described by a combination of two exponential functions, 
\begin{equation}
 \frac{A(x)}{2} = 
   \begin{cases} 
      A_1\, e^{a_1 x} & 0 \leq x \leq \ell \\
      A_2\, e^{-a_2 x} & \ell \leq x \leq L 
   \end{cases}
   \label{eq:cavity}
\end{equation}
As we are interested in asymmetrical cavities, we will use $ \ell < L/2 $. In order to maintain the continuity of the piecewise function in $x = \ell$ and $x = L$, the parameters $A_1,\,A_2,\,a_1$, $a_2,\,L,\, \ell$ are related by:
\begin{align}
  a_2  &= \frac{a_1 \ell}{(L - \ell)} \\
  A_2 &= A_1\, e^{a_2 L} \\
  a_1 & > a_2 > 0
  \end{align}

Because the particles inside the cavity are neither created nor destroyed, when the system has reached the stationary state, the current $J$ is constant, for a given force and mean concentration of particles in the cavity. With that in mind, it is possible to propose a complete general solution, which has two terms, one proportional to $ e^{(\beta F x)} $ (which corresponds to the homogeneous solution, $J = 0$), and the other proportional to $e^{-a_1 x}$ or $e^{a_2 x}$. Then we solve it for both regions defined by Eq.~(\ref{eq:cavity}).

Let us consider this expression for the concentration as a function of $x$,
\begin{equation}
  \begin{cases}
    n_1(x) = C_1\, e^{-a_1 x} + C_2\, e^{\beta F x} & 0 \leq x \leq \ell \\ 
    n_2(x) = C_3\, e^{a_2 x}  + C_4\, e^{\beta F x} & \ell \leq x \leq L 
  \end{cases}
  \label{eq:n_general}
\end{equation}
where $C_1$, $C_2$, $C_3$ and $C_4$ are constants. To determine these constants, we make use of the following conditions:

\begin{description}
 \item[$J$ is constant along the cavity] 
    If $J_1$ is the current when $x \in [0;\ell]$ and $J_2$ is the current when $x \in [\ell;L]$, then $J_1 = J_2 = J \, \forall x \in [0;L]$.
  \item[$n(x)$ is continuous]
    So we can write $n_1(0) = n_2(L)$ and $n_1(\ell) = n_2(\ell)$.
  \item[$N$ is constant]
  The total number of particles inside the cavity is constant, so the relation \mbox{$\int_0^LA(x)n(x)dx=N$} holds.
\end{description}

With these assumptions it is possible to find the exact solution to Eq.~(\ref{eq:FJ-original}) in the stationary state for the current of non--interacting particles and the concentration of particles along the cavity.
Unfortunately, the analytical expressions of the constants $C_1$, $C_2$, $C_3$, and $C_4$ of Eq.~(\ref{eq:n_general}), as functions of the parameters of the system $A_1$, $a_1$, $\ell$, and $L$ are too long to be written here. Instead, the exact analytical results of $J$ and $n$ for the particular cavity of Fig.~\ref{fig:1}, along with MC simulations will be shown in figures.

\section{Fick--Jacobs equation for interacting particles}
\label{sec:teo-conhc}

It has been shown that it is possible to find a modified version of the Fick--Jacobs equation for the case of hard-core interacting particles, see \cite{Suarez2015}.
This equation is similar to Eq.~(\ref{eq:FJ-original}), but it takes into account the fact that two particles cannot occupy the same position at the same time. This is embraced in the non-linear term $n (1-n)$:
\begin{equation}
  \frac{J}{D} = - A\frac{\partial n}{\partial x} + F A \beta n (1-n) \;\;.
  \label{eq:FJ-final}
\end{equation}

Eq.~\ref{eq:FJ-final} resumes the diffusion of many interacting particles, in a one--dimensional partial differential equation. If we have reached the stationary state, $J$ is constant, so there is no temporal dependence.
It is also important to mention that Eq.~(\ref{eq:FJ-original}) and Eq.~(\ref{eq:FJ-final}) are only valid when the concentration is almost constant in the transverse direction of the cavity, i.e., $n(x,y)\approx n(x)$.

To solve this equation, we described the boundary of the cavity and $n(x)$ with the corresponding Fourier series up to the $9^{\mathrm{th}}$ term. Finally, we obtained the coefficients of each term and found approximate expressions of the current and the concentration inside the cavity, in the stationary state. 

\begin{figure}
  \includegraphics[width=\linewidth]{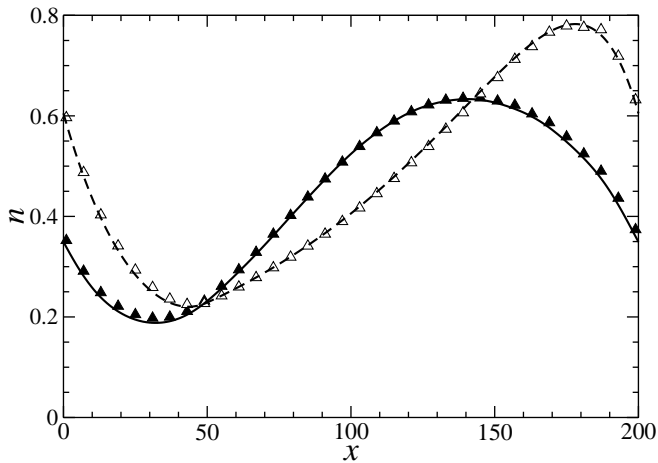}
  \caption{Concentration as a function of position when the force is applied to the right. $\delta=0.05$ and $c=0.4$. The dashed line was obtained from Eq.~(\ref{eq:FJ-original}) and $\vartriangle$ corresponds to the non--interacting particles simulation, averaged over $10^6$ configurations. The continuous line is the corresponding theoretical solutions obtained from Eq.~(\ref{eq:FJ-final}) for a system with interacting particles and {\tiny \TriangleUp} corresponds to the interacting particles simulations, averaged over $10^8$ configurations.
The time is $10^5$ MC time steps and thereafter (steady state). The length of the horizontal axis, $0\leq x \leq 200$ corresponds to the size of one cavity, see Fig.~\ref{fig:1}.}
  \label{fig:nvsx}
\end{figure}

\section{Simulations}
\label{sec:simulations}

In order to assess the validity of the results, we simulated the described system using the standard MC method.
The first step is to discretize the cavity depicted in Fig.~\ref{fig:1}. We used a two--dimensional square lattice of $200$ sites long and $50$ sites tall, with lattice spacing $a=1$. Then we draw the boundaries of the cavity inside the lattice. On the exits we set periodic boundary conditions to simulate an infinite channel of cavities. 
The process is simulated as follows. Each particle occupy only one site. A particle is selected randomly, and it is able to jump in any of the four possible directions (up, down, left or right).

\begin{figure}
    \includegraphics[width=\linewidth]{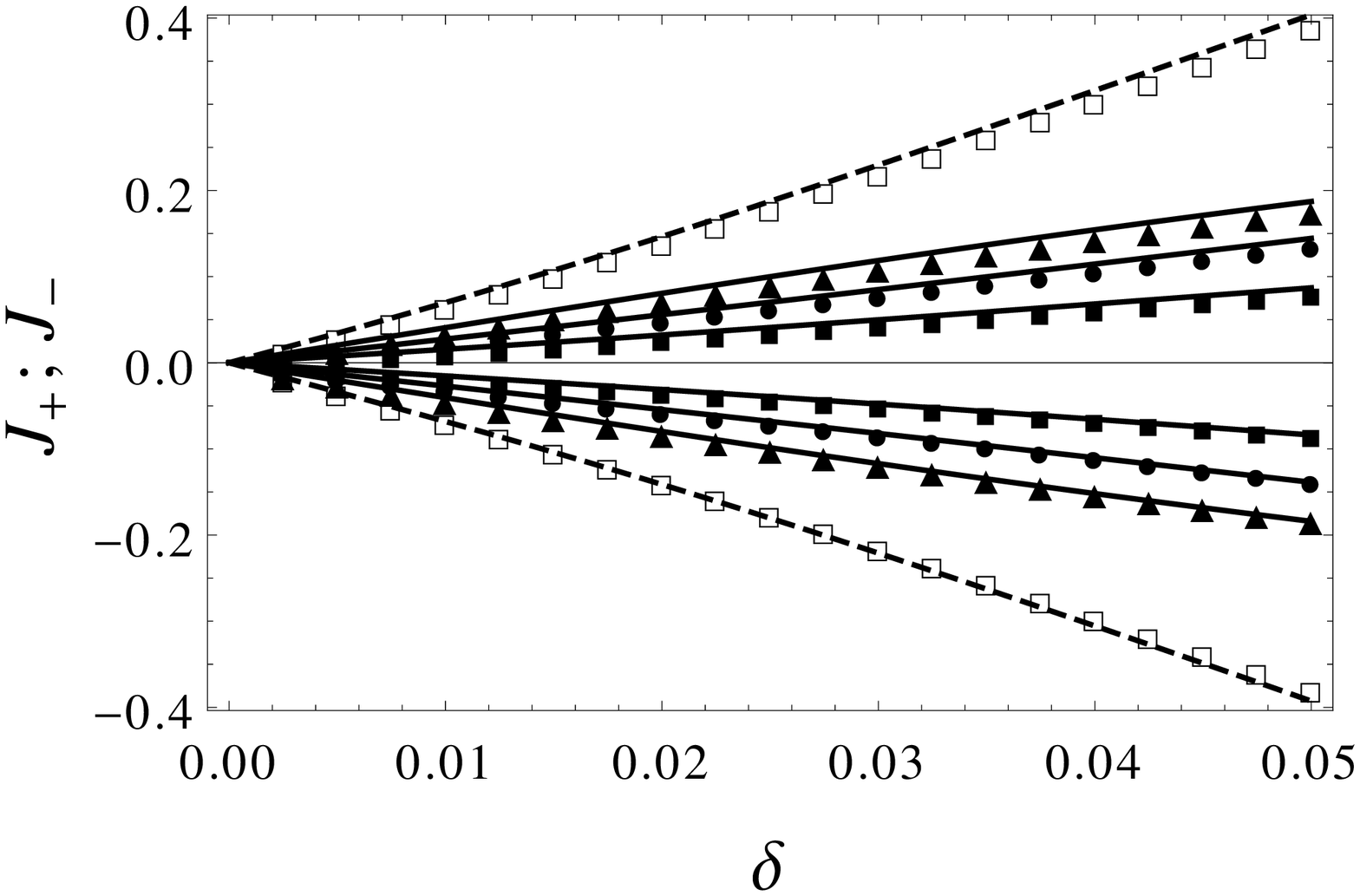}
    \caption{
      \label{fig:jotas} Current of particles when the force is applied to the left ($J_- < 0$) or to the right ($J_+ > 0$). The continuous lines are the corresponding theoretical solutions obtained from Eq.~(\ref{eq:FJ-final}). The dots correspond to numerical simulations. {\tiny \SquareSolid}: $c=0.1$. {\tiny \CircleSolid}: $c=0.2$. {\tiny \TriangleUp}: $c=0.4$. The dashed line is the solution to Eq.~(\ref{eq:FJ-original}), and {\tiny \Square} are the corresponding simulation with non--interacting particles for $c=0.4$. In all cases we perform averages over $2 \,  10^6$ configurations. The time is $10^5$ MC time steps.
    } 
\end{figure}

The jumping rate of each direction is determined by the applied force. For example, if there is no force, all the particles have the same jumping rate in any direction. The jumps have a length equal to the size of a particle.
If the force is applied to the right in the $x$--axis, the jump rate is:
\begin{align}
  \begin{split}
    P_{\rightarrow} &= p (1 + \delta) \\
    P_{\leftarrow} = P_{\uparrow} = P_{\downarrow } &= p
  \end{split}
\end{align}
where $\delta = \beta |F| a$, and we set $p = 1$ for simplicity.  Analogously, if the force is applied to the left,
\begin{align}
  \begin{split}
    P_{\leftarrow} &= p (1 + \delta) \\
    P_{\rightarrow} = P_{\uparrow} = P_{\downarrow } &= p
  \end{split}
\end{align}
The relation between the dimensionless external force $\delta$ and the actual force $F$ acting on the particle is given by $1 + \delta = e^{\beta |F| a}$.
Considering $\delta \ll 1$, it is possible to assume \mbox{$\delta \approx \beta |F| a$.}

If a particle tries to move into an occupied site, the process will not take place. That is how we represent the hard--core interaction. In other words, it is only one particle allowed on each site of the lattice. If the particles do not interact with each other, this condition is not fulfilled; i.e. multiple occupation is allowed. In all cases the jumps that cross the wall of the cavity are forbidden.
Another key question that has to be considered is the temporal discretization. We say that one time step has passed when, in average, every particle had the opportunity to jump once. That is a convention widely used in MC simulations.

The mean concentration $c$ in the cavity is given by,
\begin{equation}
    c = \frac{N}{\int_0^{L} A(x)dx} \; ,
\end{equation}
where $N$ is the total number of particles inside the cavity, and in the discrete case, 
\mbox{$\int_0^{L} A(x)dx$} is the total number of sites inside the cavity.
Once all the rules have been set up, we let the system evolve until the stationary state has been reached. After that, we can measure any relevant quantity. In this paper we will consider the current of particles that cross the cavity per unit of time, $J$, the concentration, $n$, and the mobility, $\mu$. 

\begin{figure}
    \includegraphics[width=\linewidth]{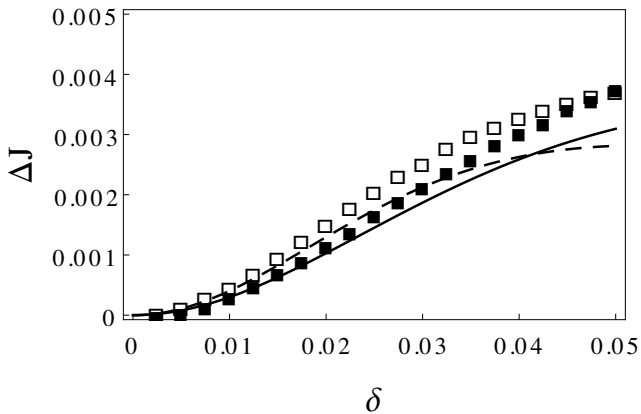}
      \caption{
      \label{fig:deltaJ-0.1} Difference in the current as a function of the force. The mean concentration is $c=0.1$. The dashed line was obtained from Eq.~(\ref{eq:FJ-original}) and {\tiny \Square} corresponds to the non--interacting particles simulation. The continuous line is the corresponding theoretical solution obtained from Eq.~(\ref{eq:FJ-final}) and {\tiny \SquareSolid} corresponds to the hard--core interacting particles simulation. We perform averages over at least $2.5 \,  10^6$ configurations and the time is $10^5$ MC time steps in both cases.} 
\end{figure}

\section{Results and discussion}
\label{sec:results} 

We solved Eq.~(\ref{eq:FJ-final}) for the same cavity described in Section~\ref{sec:teo-sinhc} considering particles that interact through hard--core potential. 
The first quantity we can measure to check the validity of Eq.~(\ref{eq:FJ-final}) is the concentration profile, $n(x)$. The results for interacting and non--interacting particles are shown in Fig.~\ref{fig:nvsx}.
It can be appreciated that both theoretical solutions match the simulation points. 
Although they are similar in some sense, it is clear that the interaction plays an important role in the way particles arrange inside the cavity. It shifts horizontally the location of the maximum of concentration $n(x)$ and makes it less pronounced. This effect is larger when the force increases. A more detailed discussion of this result can be seen in \cite{Suarez2015}.

\begin{figure}[t]
    \includegraphics[width=\linewidth]{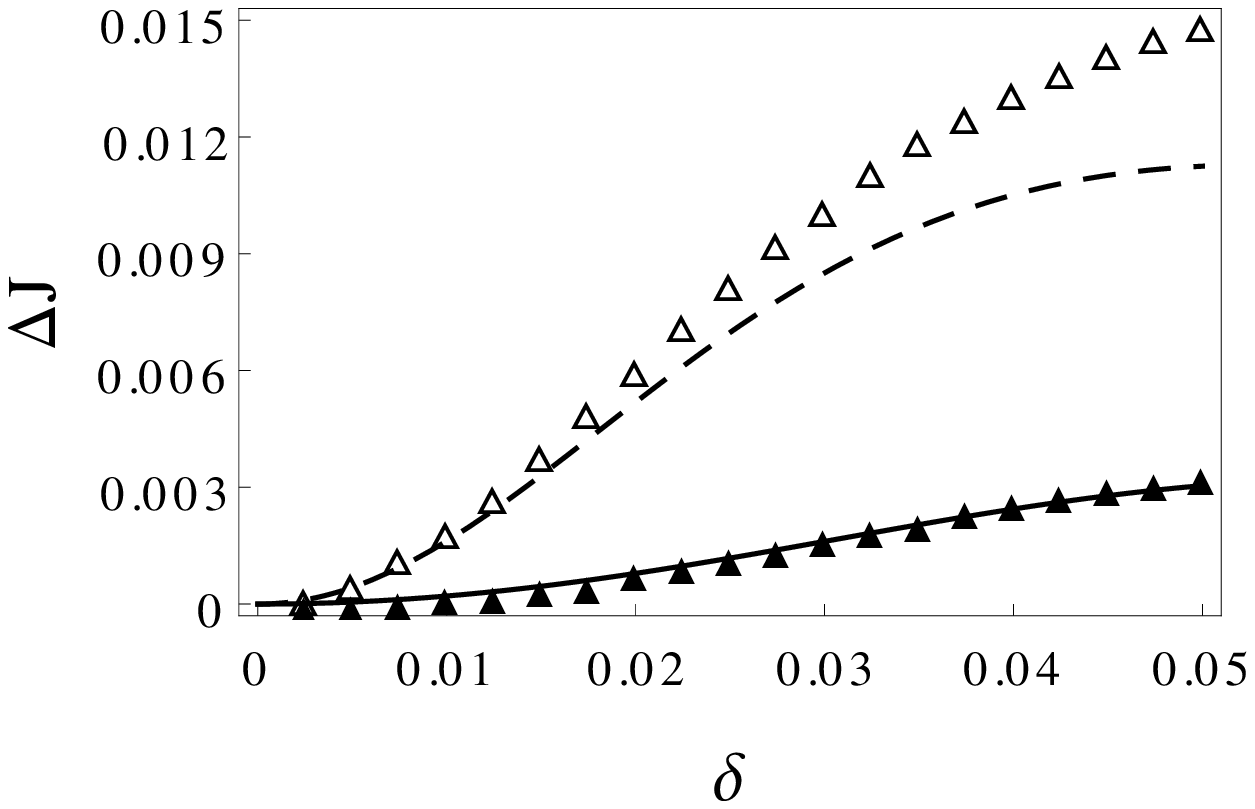}
      \caption{\label{fig:deltaJ-0.4} Difference in the current as a function of the force. The mean concentration is $c=0.4$. Dashed line was obtained from Eq.~(\ref{eq:FJ-original}) and  $\vartriangle$ corresponds to the non--interacting particles simulation. The continuous line is the corresponding theoretical solution obtained from Eq.~(\ref{eq:FJ-final}) and {\tiny \TriangleUp } corresponds to the hard--core interacting particles simulation. We perform averages over $5 \,  10^6$ configurations and the time is $10^5$ MC time steps in both cases.}  
\end{figure}

Furthermore, in Fig.~\ref{fig:jotas} we can see the current obtained when the force is applied to the left ($J_- < 0$) and when the force is applied to the right ($J_+ > 0$) for different values of the mean concentration. We also show in Fig.~\ref{fig:deltaJ-0.1} and Fig.~\ref{fig:deltaJ-0.4}  the difference of these two quantities, $\Delta J = J_+ + J_- = |J_+| - |J_-|$. Due to the asymmetry of the cavity, $\Delta J  > 0$ for the non interacting system as expected (see, e.g. Ref.~\cite{Reguera2012}).

We compared the results obtained for non-interacting and hard-core interacting particles in Fig.~\ref{fig:deltaJ-0.1} and Fig.~\ref{fig:deltaJ-0.4}. The solution to Eq.~(\ref{eq:FJ-final}) is expressed as an infinite expansion of harmonic functions, that in fact has been truncated up to the $9^{th}$ term. Instead, the solution to the non--interacting case is the exact solution to Eq.~(\ref{eq:FJ-original}).
Considering these two plots, we can see that for the non-interacting case the only effect of the change in the mean concentration is, as expected, a rescaling of the vertical axis. Analytical results depart from numerical simulations as the force is increased, since larger forces imply a worsening of the conditions for the validity of the Fick--Jacobs equation. Nevertheless, we wish to highlight that the hard-core interaction enlarges the validity range of the Fick--Jacobs equation, as can be appreciated in Fig.~\ref{fig:deltaJ-0.4}, where the concentration is higher.

Another quantity used to describe the transport of particles inside the cavity is the mobility, $\mu$, defined as, 
\begin{equation}
  \mu_\pm = \frac{\left| v_{d_\pm} \right|}{\delta}
  \label{eq:mu}
\end{equation} 
where $v_{d_+}$ ($v_{d_-}$) is the mean drift velocity when the force is applied to the right (left).

The average time $t$ that a particle needs to cross a whole cavity is $t_\pm=L/|v_{d_\pm}|$. After this amount of time has passed, all the $N$ particles also have crossed the cavity.
Then \mbox{$|J_\pm| = N/t_\pm $}, and substituting in Eq.~(\ref{eq:mu}) one has,
\begin{equation}
\left| v_{d_\pm} \right| = \frac{\left| J_{\pm} \right| L}{N \delta}\;.
\end{equation}

\begin{figure}
  \includegraphics[width=\linewidth]{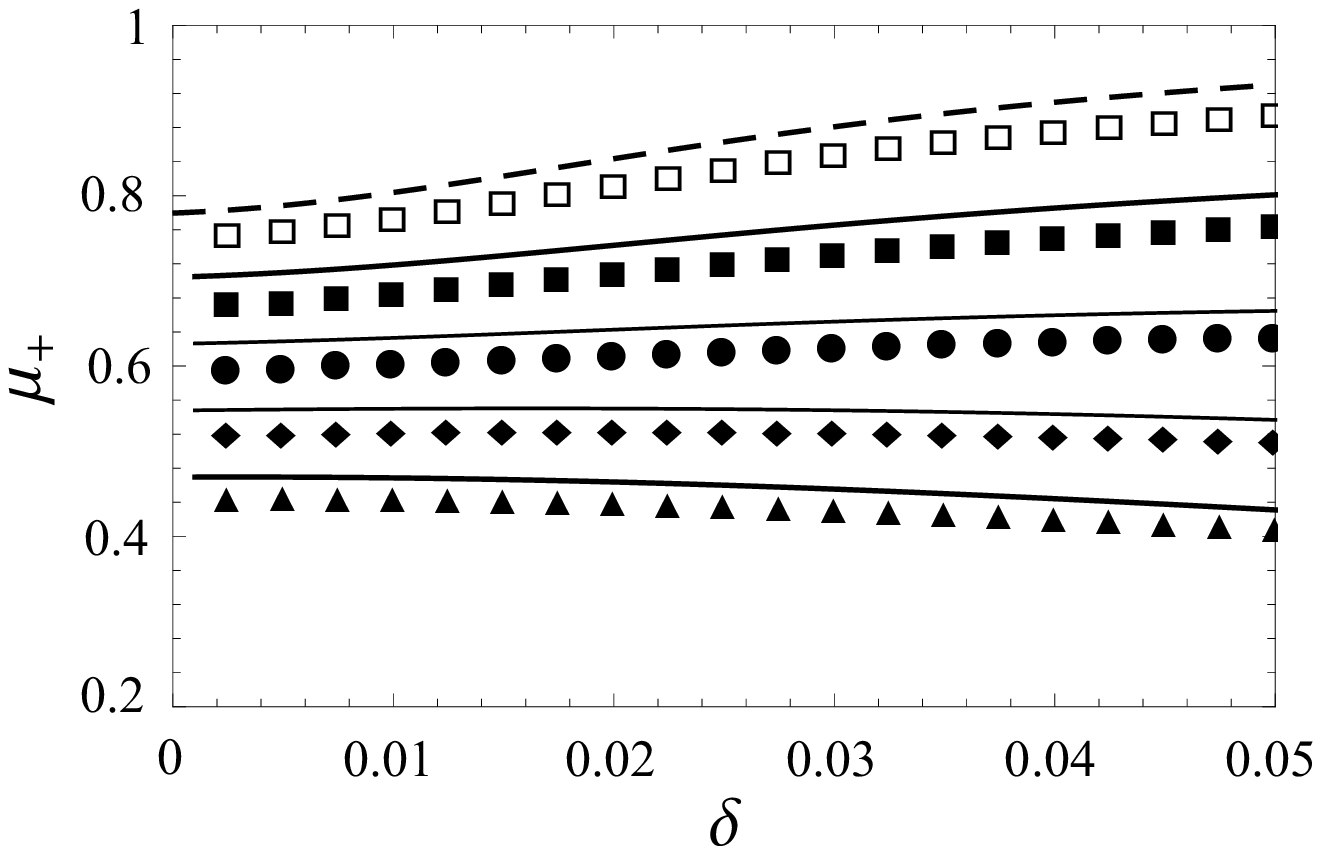}
  \caption{Mobility as a function of the force when the force is applied to the right. The dashed line was obtained from Eq.~(\ref{eq:FJ-original}) and {\tiny \Square} corresponds to the non--interacting particles simulation. The continuous lines are the corresponding theoretical solutions obtained from Eq.~(\ref{eq:FJ-final}) for each mean concentration: $c=0.1$ ({\tiny \SquareSolid}), $c=0.2$ ({\tiny \CircleSolid}), $c=0.3$ ({\tiny \DiamondSolid}) and $c=0.4$({\tiny \TriangleUp}).
  In all cases we perform averages over $2\, 10^6$ configurations. The time is $10^5$ MC time steps.} 
  \label{fig:mu-mas} 
\end{figure}

\begin{figure}[th]
  \includegraphics[width=\linewidth]{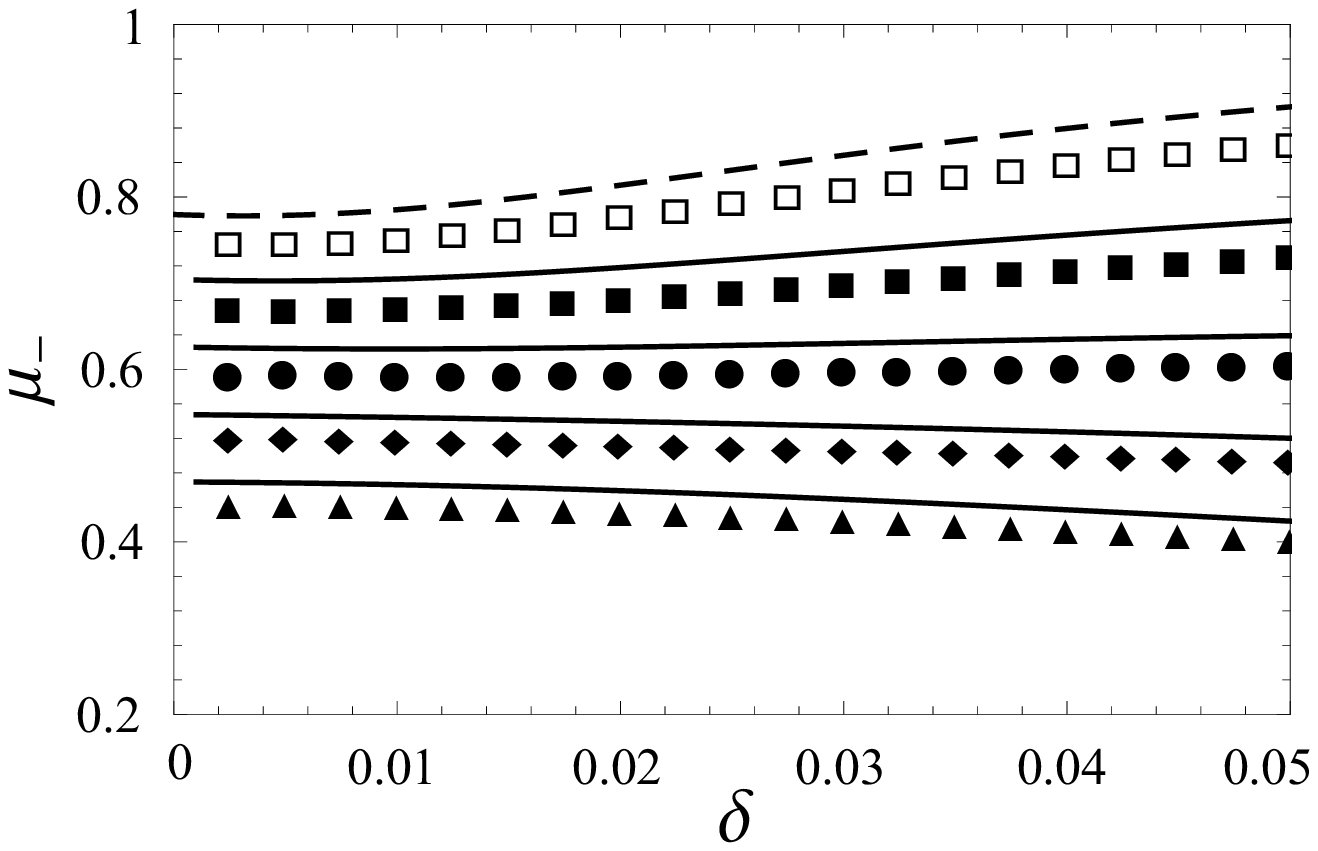}
  \caption{Mobility as a function of the force when the force is applied to the left. The dashed line was obtained from Eq.~(\ref{eq:FJ-original}) and {\tiny \Square} corresponds to the non--interacting particles simulation. The continuous lines are the corresponding theoretical solutions obtained from Eq.~(\ref{eq:FJ-final}) for each mean concentration: $c=0.1$ ({\tiny \SquareSolid}), $c=0.2$ ({\tiny \CircleSolid}), $c=0.3$ ({\tiny \DiamondSolid}) and $c=0.4$({\tiny \TriangleUp}).
  In all cases we perform averages over $2\, 10^6$ configurations. The time is $10^5$ MC time steps.}
  \label{fig:mu-menos}
\end{figure}

In Figs.~\ref{fig:mu-mas} and~\ref{fig:mu-menos} we can see the mobility as a function of the force applied for positive and negative force and for different concentrations.
For a given value of $\delta$, the highest mobility corresponds to the non--interacting case, and as we increase the concentration, with hard--core interaction, the mobility reduces.
As expected for a non--interacting system the mobility $\mu_+$ and $\mu_-$ increases with $\delta$ (see e.g. \cite{Dagdug2012}). This is also the behavior of $\mu_+$ and $\mu_-$ for interacting systems when the mean concentration $c$ is very small. For larger values of $c$ the dimensionless force $\delta$ produces the accumulation of particles near the exit of the cavity where the transverse area $A$ is small. This effect increases with $\delta$, and the mobility is reduced (see Figs.~\ref{fig:mu-mas} and~\ref{fig:mu-menos}).

Due to the particle--hole symmetry, for a given dimensionless force $\delta$, it is possible to obtain (independently of the cavity shape, see \cite{Suarez2013}) that
\begin{equation}
  (1-c) \Delta J^{1-c} = - c \Delta J^c
  \label{eq:ph}
\end{equation}
where $\Delta J^c=|J^c_+|-|J^c_-|$ is the difference in the current for a given mean concentration $c$. This condition is trivially fulfilled in the MC simulation (when a particle jumps to the right, one hole jumps to the left, or vice versa), but it is also obtained form Eq.~(\ref{eq:FJ-final}) (see \cite{Suarez2015}). Eq.~(\ref{eq:ph}) implies that $\Delta J=0$ for $c=0.5$. This is in agreement with the comparatively small value of $\Delta J$ obtained for $c=0.4$ in Fig.~\ref{fig:deltaJ-0.4}.

It is important to notice that, in all cases, the analytic solution for the mobility is slightly higher than the simulations, even for the exact solution of the non--interacting case. This is probably a consequence of the discretization of the system in the simulation, that is not present in the Fick--Jacobs equation.
%
%
\section{Conclusion}
\label{sec:conclusion}

We considered the transport of particles inside a chain of cavities with exponential shape. We compared the current, the concentration and the mobility of interacting and non--interacting particles. The former are described by the Fick--Jacobs equation; in this particular case, we found the exact solution. The latter is described by a similar equation with a non--linear term, that takes into account the hard--core interaction between particles. In this case the solution is expressed by a finite sum of harmonic functions. 

We can conclude that, when the concentration is low, both systems behave in a similar manner. The interaction does not play an important role and the non--linear term in Eq.~(\ref{eq:FJ-final}) is negligible. In contrast, for higher concentrations, the non--linear term becomes significant. This implies that, when the concentration is high enough, the current of particles is mainly determined by the interaction between particles and not so much by the interaction of the particles with the borders of the cavity.

For interacting particles one can see that even when there is a small difference between the results of $J$ and $\mu$ obtained from the generalization of the Fick--Jacobs equation from the ones obtained from MC simulations for the same system, the general tendencies of all the analytical results as a function of concentration and external force are verified.

Comparing the analytical results with the corresponding MC results, one can note that the results of the mobility $\mu$, obtained from the modified Fick--Jacobs equation for interacting particles, are at least as good as the ones obtained from the original Fick--Jacobs equation for non-interacting particles. For the current difference $\Delta J$ one finds a better agreement for the case of interacting particles.
On the other hand, a very good agreement between the concentration $n(x)$ obtained from the modified Fick--Jacobs equation and numerical results was found. In summary, the generalization of the Fick--Jacobs equation gives a good description for the diffusion and the transport of particles with hard-core interaction in narrow channels.


%

\end{document}